\documentclass[twoside,11pt]{article}

\usepackage{cjp}
\usepackage{graphics}

\begin{document}
 \title{Thermoelectric effects in superconductors}
\author{Jan Kol\'a\v cek$^1$, 
        Tzong-Jer Yang$^2$}
\address{$^1$Institute of Physics, ASCR, Cukrovarnick\'a 10, 
         16253 Prague 6, Czech Republic\\ 
        $^2$Dept. of Electrophysics,National Chiao-Tung
        University, Hsinchu 30050, Taiwan, ROC}
     
\date{Jul. 8, 2004}
\maketitle              
\begin{abstract}
It is widely believed that temperature gradient does not induce 
electric field in the superconductor and consequently that 
thermoelectric effects do not exist, or are negligible 
in these materials. This statement is correct only as far as 
effective electric field or gradient of the electrochemical 
potential is concerned. In normal metals temperature gradient 
generates effective electric field, which nulls out thermally induced 
diffusion current. In superconductor the diffusion current of 
quasiparticles is canceled a counterflow of supercurrent. Superconducting
current induces the true electric field, which can be approximated by gradient 
of the screened Bernoulli potential. It explains familiar 
giant thermomagnetic flux observed in superconducting thermocouples.
Contactless measurements of thermoelectric effects are suggested.
\end{abstract}


\section{Introduction}
In the normal metal temperature gradient induces electric field, which 
nulls out thermally induced diffusion current. If two different metals 
are connected together so that they form a closed loop thermocouple, 
temperature gradient induces circulating thermocurrent. 
In superconductors the diffusion current of quasiparticles 
$j_n=-L_T \nabla T $ is canceled by a counterflow of supercurrent.
Measured thermoelectric voltage is zero or negligibly small what is
the base for widely spread opinion, that thermoelectric effects do not
exist in superconductors. Nevertheless, it was experimentally proved 
that in the superconducting thermocouple temperature gradient induces 
giant flux \cite{80VanHarlingen}, proportional to the transport
coefficient $L_T$ measured above $T_c$. 
The question is: what is the origin of the current which evocates the 
emerging flux?
 
\section{The true and the effective electric field}
The answer to this question is based on a simple notion. 
The true electric field properly used in Maxwell equations must be 
distinguished from the effective electric field defined as a 
gradient of electrochemical potential. At equilibrium the
electrochemical potential is constant and it implies that 
gradient of the electrochemical potential, often refferred to as 
the effective electric field, is zero.
Voltmeter measures difference of the 
electrochemical potential between the two contacts and by a lot of 
measurements the zero (effective) electric field in superconductor 
was unambiguously proved. But it does not say much about the
true electric field. Already in the very early theories 
it was predicted (see e.g. \cite{50London}), that 
superconducting current generates perpendicular (true) electric
field. The prediction was experimentally verified. By a
contactless capacitive pickup the presence of nonzero (true) Hall 
voltage in the superconductor in Meissner state was proved 
(see e.g. \cite{68BokKlein}). From the measurements, 
from the BCS theory \cite{75Hong} and also from the
Ginzburg-Landau theory \cite{KL01,LKMB02} 
it follows, that the true electric field can be approximated by 
the gradient of the screened Bernoulli potential
\begin{equation} 
e\varphi\approx n_s/n\left(-{1\over 2}m{\bf v}_s^2\right),
\end{equation}
where ${\bf v}_s$ denotes velocity of the superconducting particles,
$n_s$ and $m$ their density and mass. 

\section{Thermoelectric effects in superconductors}
\begin{figure}[h]
\centerline{\includegraphics{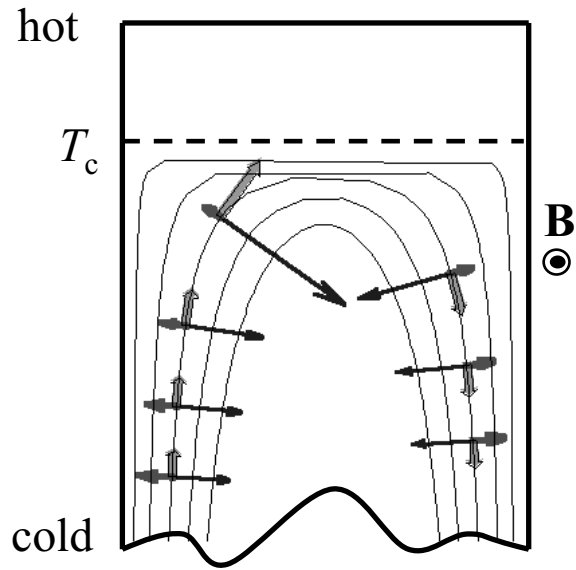}}
\figcaption{Schematic picture of the screening current 
in a superconducting slab with temperature gradient.}
\label{ScreeningCurrent}
\end{figure}

In superconductors the effective electric field generated by 
the temperature gradient is zero, or negligibly small.
Here we show that the true electric field induced by the temperature 
gradient is relevant. Let us imagine 
a single superconducting slab $-d<x<d$ in external magnetic field $B_0$,
directed parallel to the $z$ axis. Superconductor in Meissner state 
exponentially screens magnetic field, so that 
$B=B_0{\cosh(x/\lambda) / \cosh(d/\lambda)}$.
If temperature and consequently also the London penetration depth 
$\lambda=\sqrt {m/\mu_0 n_s(T) e^2}$ increases along the $y$ axis, 
from Maxwell equation  
$\mu_0 {\bf j}=\nabla \times {\bf B}
               =\left[\partial_y B,-\partial_x B,0  \right]$ 
follows that screening current has nonzero component also 
in the $x$ direction. The trajectories of superconducting particles 
are  schematically sketched on 
Fig.\ref{ScreeningCurrent}. 
 
On the cold end (foot of the picture) the penetration depth is
small and screening current is concentrated close to the surface. 
On the side where screening current flows in the direction of 
increasing temperature (left hand side on the picture), 
the velocity of the Cooper pairs (marked by double arrows) increases, 
near $T_c$ changes direction and returns back on the other side. 
Lorentz and electric field forces marked by
thin and thick arrows, respectively, are perpendicular to the velocity of 
superconducting particles. Note, that the electric field force does not
fully compensate the Lorentz force. The balance of forces is assured
by the superconducting-normal state fluid (s-n) interaction, 
in the literature often called quasiparticle screening \cite{VS64}. 
  
\begin{figure}[h]
\centerline{\includegraphics{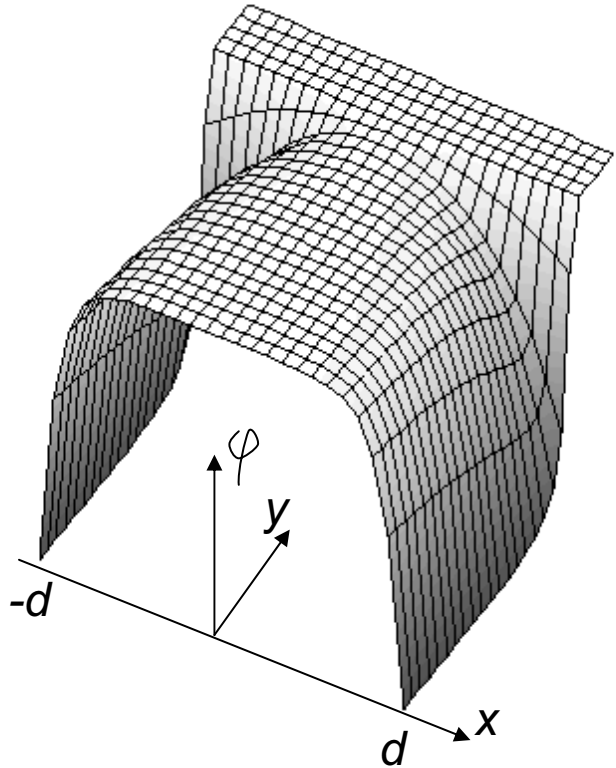}}
\figcaption{Screened Bernoulli potential in superconductor. 
At position where temperature 
reaches $T_c$ a sharp, but continuous step is present.}
\label{BernoulliPot}
\end{figure}
The scalar potential in the screened Bernoulli approximation 
is sketched in the figure \ref{BernoulliPot}. 
Let us note that as expected, the sharp step at position where temperature
reaches $T_c$ is continuous - in the limit 
$t \rightarrow 1$ the scalar potential $\varphi \rightarrow 0$. 
We must stress, that presented drawing displays scalar potential, which is not 
measurable by standard voltmeter measurement with contacts, but which
should be observable by contactless measurements.
  
\section{Summary}
In ref. \cite{04thermopower} it was showed, that taking into account 
the electrostatic potential it is possible to treat thermoelectric
properties in superconductors on the same footing as in the normal
metals. From the scalar potential matching thermally induced
magnetic flux in agreement with experiment was found. Nevertheless, 
one open question remained. In the cited paper we have matched
Bernoulli potential only at the inner surface of the thermocouple. 
By showing here the distribution of the scalar potential in the 
close vicinity of $T_c$, brought arguments indicating, 
that the solution has the desired properties 
in the whole cross-section. Last but not least, we would like to encourage  
contactless measurement of the thermoelectric effects in superconductors.  

\section*{Acknowledgment}
This work was supported by MSMT program Kontakt ME601 and GACR
202/03/0410, GAAV A1010312 grants. One of us (JK) acknowledges 
also support of National Science Council, Taiwan, Republic of China. 


\end{document}